\documentclass[aip,apl,reprint,superscriptaddress]{revtex4-2}
\usepackage{gensymb}
\usepackage{graphicx}
\usepackage{amsmath, amsthm, amssymb, amsfonts}
\usepackage{hyperref}
\usepackage{xcolor}
\usepackage{xcolor}
\usepackage{array}
\usepackage{booktabs}
\usepackage{listings}
\usepackage{tikz}

\begin{document}

\title{Disorder-Induced Griffiths like Phase and Charge Localization in Sr$_2$FeRuO$_6$}

\author{Koustav Pal}
\email{koustavpal001@scut.edu.cn}
\email{koustav.pal97@gmail.com}
\affiliation{Spin-X Institute, School of Physics and Optoelectronics, State Key Laboratory of Luminescent Materials and Devices, Guangdong-Hong Kong-Macao Joint Laboratory of Optoelectronic and Magnetic Functional Materials, South China University of Technology, Guangzhou 511442, China}
\affiliation{Institute of Engineering and Management, Sector-V, Kolkata 700091, West Bengal, India}
\affiliation{University of Engineering and Management, Newtown, Kolkata 700160, West Bengal, India}
\affiliation{Saha Institute of Nuclear Physics, A CI of Homi Bhabha National Institute, 700064, Kolkata , India}

\author{Shuvendu Ghosh}
\affiliation{Department of Physics, Aditya University, Surampalem, 533437, Andhra Pradesh, India}	

\author{I. Das}
\affiliation{Saha Institute of Nuclear Physics, A CI of Homi Bhabha National Institute, 700064, Kolkata , India}

\date{\today}

\begin{abstract}

\begin{center}
	\textbf{Abstract}
\end{center}

A comprehensive understanding of complex transition-metal oxides is key to designing next-generation spintronic materials with tailored magnetic and transport functionalities. We investigate the interplay between structural disorder, frustrated magnetism, and localized charge transport in the 3d–4d hybrid double perovskite {Sr$_2$FeRuO$_6$}. The compound adopts a monoclinic structure with significant B-site antisite disorder that drives its unconventional behavior. Magnetic measurements uncover a robust Griffiths phase extending well above the ordering temperature, evidencing stable short-range ferromagnetic clusters within a disordered matrix. Electrical transport remains strongly localized, transitioning from {Small-Polaron Hopping} at high temperatures to {Variable-Range Hopping} at intermediate temperatures. 
{\color{black}
	These results position Sr$_2$FeRuO$_6$ as a representative 3d$-$4d hybrid system in which quenched disorder stabilizes spin inhomogeneity, while polaronic and hopping transport are also observed.
}

\end{abstract}


	\maketitle

\section{Introduction}

In recent years, the rapid advancement of science and technology has intensified the demand for magnetic materials exhibiting tailored transport phenomena \cite{Khan2022, Anik2021, Han2023, Pal2024}. Such materials are indispensable across a wide range of applications—from aerospace and communication technologies to biomedical devices—where efficiency, reliability, and multifunctionality are critical. Consequently, a fundamental understanding of magnetic and transport phenomena is essential for the development of next-generation energy-efficient and high-performance devices. In this pursuit, various material classes, including intermetallic compounds \cite{Koushik2024, Ghosh2023}, transition-metal oxides, and two-dimensional systems, have been extensively investigated in different physical forms such as bulk, thin films, and nanostructures \cite{Trinadh2023, Krishnaiah2025, 9, 10, 11}.

Among these, perovskite oxides have attracted considerable attention owing to their structural flexibility and tunability. The generic perovskite structure, $ABO_3$, can host a wide variety of cations at both the A and B sites, enabling the fine-tuning of electronic bandwidth, lattice distortion, and exchange interactions. An extended variant of this structure—the double perovskite with general formula $A_2B'B''O_6$—introduces an additional degree of freedom by allowing two distinct transition-metal ions at the B site \cite{4,5,6,PAL2024120276,PAL2024140211}. This cationic diversification gives rise to a rich array of physical properties, from magnetoresistance and multiferroicity to spin-polarized transport. The magnetic and transport behavior of double perovskites is strongly influenced by the ordering of the B-site cations, where antisite disorder (ASD) the random interchange of $B'$ and $B''$ ions acts as a fundamental control parameter \cite{7,8}.

The properties of these compounds become particularly interesting when involving an intermixing of 3d, 4d, and 5d elements, as this creates a large variation in local environments, ranging from localized orbital behavior to strong spin-orbit coupling. Hybridization between localized 3d orbitals and more delocalized 4d/5d states within the $A_2B'B''O_6$ lattice results in a complex interplay between exchange interactions, electronic correlations, and spin–orbit coupling \cite{Pal2024,  Pal20246h}. These competing factors have led to the emergence of remarkable functional properties such as half-metallicity and high Curie temperatures ($T_C$), making such systems promising candidates for spintronic applications \cite{9,10,11,PAL2023170794}. Benchmark double perovskites, including Sr$_2$FeMoO$_6$ and Sr$_2$FeReO$_6$, exemplify how ASD crucially determines both the magnetic moment and $T_C$ \cite{12,13,14,15,16,17,18}. Therefore, understanding the role of structural disorder is indispensable for elucidating the magnetic behavior of 3d–4d/5d hybrid systems.

A particularly intriguing member of this family is Sr$_2$FeRuO$_6$ (SFRO), which combines the magnetic characteristics of 3d–Fe and 4d–Ru ions. The inherent antiferromagnetic nature of SrFeO$_3$ and the ferromagnetic behavior of SrRuO$_3$ together establish a delicate balance of competing interactions in SFRO \cite{s1,s2,s3}. Previous studies have demonstrated that this competition, in conjunction with ASD, gives rise to a spin-glass ground state \cite{831,mtc}, wherein the random distribution of Fe and Ru ions produces a frustrated network of ferromagnetic (FM) and antiferromagnetic (AFM) exchange interactions. This frustration inhibits the establishment of long-range magnetic order, leading instead to a frozen disordered state at low temperatures \cite{r2,r3,Ghosh2025Ev,12,13,KARIKAR2026114021,AHMED2025114760}.

Such disorder-driven systems often exhibit unconventional magnetic transitions. The quenched ASD, being a static lattice feature, can influence not only the ground state but also the paramagnetic (PM) regime. Notably, the transition from the PM phase can become smeared, reflecting the formation of short-range magnetic clusters well above the nominal ordering temperature. This phenomenon is described by the Griffiths phase (GP), wherein locally ordered FM clusters persist within a globally PM matrix in the temperature range between the Griffiths temperature ($T_G$) and $T_C$ \cite{Saha2022, Ghosh2025}. The appearance of the GP in SFRO is thus a natural extension of the same competing interactions and disorder mechanisms that generate its spin-glass behavior.

Another important aspect linked with these type of material of having transport behaviour. The high electrical resistivity of this compound poses significant experimental challenges, often rendering conventional measurement techniques ineffective. In the present study, we have carefully investigated the transport properties of SFRO using low-current measurements with optimized noise control. Our analysis reveals evidence of multiple hopping mechanisms governing charge transport, alongside short-range magnetic correlations persisting well above the spin-freezing temperature. These findings provide deeper insight into the coupling between magnetic disorder and transport phenomena in 3d–4d hybrid double perovskites.

	\section{Experimental Details}

\begin{figure*}
	\centering
	\includegraphics[width=.95\linewidth]{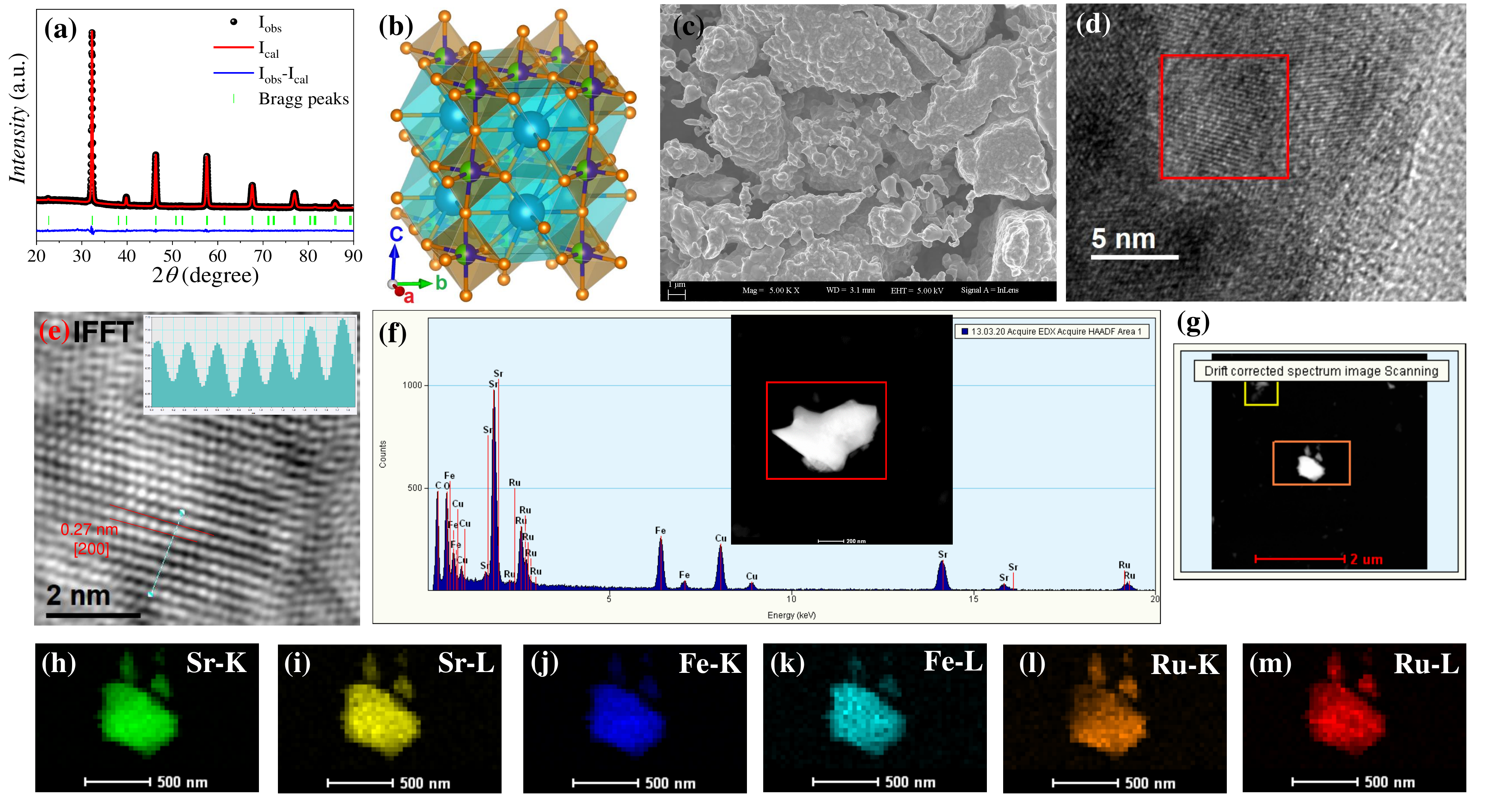}
	\caption{(a) Full-profile Rietveld refinement of the XRD pattern for the powdered polycrystalline SFRO compound recorded at 300 K; (b) Monoclinic crystal structure of the double perovskite SFRO, where FeO$_6$ and RuO$_6$ octahedra are illustrated with Fe (green) and Ru (blue) atoms at the centers surrounded by oxygen atoms (orange). The central atom in the structure is Sr (sky blue); (c) SEM image of SFRO; (d) HRTEM image of SFRO; (e) Inverse fast Fourier transform (IFFT) image obtained from the red boxed region in (d); (f) EDX spectrum of SFRO, with the inset showing the corresponding STEM-HAADF image of the analyzed particle (red box); (g) STEM-HAADF image of the particle on which elemental mapping was performed (orange box); (h)–(m) Elemental maps of Sr, Fe, and Ru.}
	\label{str}
\end{figure*}

Polycrystalline samples of SFRO were synthesized via the conventional solid-state reaction route. High-purity (99.995\%) precursors—SrCO$_3$, Fe$_2$O$_3$, and RuO$_2$—were weighed in stoichiometric proportions, thoroughly mixed, and subjected to multiple cycles of grinding and heat treatment. The initial calcination was carried out at 950$^\circ$C for 16~h, followed by a final sintering step at 1150$^\circ$C for 24~h to ensure phase formation and homogeneity.  

Structural characterization of the sintered powders was performed using X-ray diffraction (XRD) on a Rigaku diffractometer equipped with Cu-K$\alpha$ radiation ($\lambda$ = 1.5406~\AA). Diffraction data were collected over a 2$\theta$ range of 20$\degree$–90$\degree$. The crystal structure was analyzed through Rietveld refinement of the XRD data using the \textsc{FullProf} software package \cite{f1,f2}. A Pseudo-Voigt function was employed to model the peak profiles, effectively accounting for both Gaussian and Lorentzian contributions. Parameters related to instrument were optimized during refinement, allowing for precise determination of lattice parameters and the unambiguous identification of the crystallographic space group.  

For Transmission Electron Microscopy (TEM), the polycrystalline powder was finely ground and ultrasonicated in ethanol to obtain a uniform dispersion of small, thin particles. A few drops of the suspension were deposited onto copper grids for microscopic examination. TEM, High-Resolution TEM (HRTEM), Energy-Dispersive X-ray Spectroscopy (EDX), and elemental line profiling were carried out using an FEI Tecnai G$^2$ F30-ST microscope operating at 300~kV. The instrument was equipped with a Fischione model~3000 high-angle annular dark-field (HAADF) detector to obtain images free from Bragg diffraction effects, as well as an EDX system for compositional analysis \cite{dfa,gfa}.  
Grain morphology and size distribution were further examined using a Zeiss Supra~40 scanning electron microscope (SEM).  

Magnetic measurements were performed using a Quantum Design SQUID magnetometer. Both isothermal magnetization [$M(H)$] and temperature-dependent magnetization [$M(T)$] were measured to probe the magnetic behavior. 
{\color{black}
Electrical transport measurements were carried out in a cryogenic setup using a Keithley~6251A sourcemeter in the standard four-probe configuration. For the transport measurement, a rectangular pellet with approximate dimensions of \(5.57 \times 2.27 \times 0.76~\mathrm{mm}^3\) was used. Low excitation currents were employed to minimize thermal effects, and special care was taken to optimize noise filtering in view of the high resistivity of the sample. These precautions ensured reliable resistivity data acquisition under near-ohmic measurement conditions.}

	\section{Results and Discussions}
		
	\subsection{Structural properties}
	
Figure~\ref{str}(a) shows the X-ray diffraction (XRD) pattern of Sr$_2$FeRuO$_6$ (SFRO) along with the corresponding Rietveld refinement profile. Figure~\ref{str}(b) illustrates the schematic representation of the perovskite unit cell of the compound. The refinement confirms that SFRO crystallizes in a monoclinic structure with space group \textit{I2/c} (No.~15)~\cite{831}. The absence of any secondary diffraction peaks further attests to the high phase purity and single-phase nature of the synthesized material.  

{\color{black}
The refined lattice parameters were determined to be 
$a = 5.525(2)$~\AA, $b = 5.536(1)$~\AA, $c = 7.842(3)$~\AA, and 
$\beta = 90.181(1)^\circ$, with an average isotropic thermal displacement parameter of 0.01~\AA$^2$. 
In the monoclinic lattice, Fe and Ru ions occupy the corner sites represented by blue and green spheres, respectively, 
while Sr ions, shown in sky blue, reside at the central positions of the unit cell.  

\begin{table}[htbp]
	\centering
	\caption{Lattice parameters and atomic positions of cations and anions obtained from full Rietveld refinement of XRD data collected at room temperature. The reliability factors are $\chi^2 = 1.2$, $R_p (\%) = 2.25$, $R_{wp} (\%) = 2.75$, $R_{exp} (\%) = 1.67$, and the Goodness of Fit (GoF) = 1.09. The lattice parameters are $a = 5.525(2)$~\AA, $b = 5.536(1)$~\AA, $c = 7.842(3)$~\AA, and $\beta = 90.181(1)^\circ$}
	
	\begin{tabular}{l l c c c c}
		\hline
		\textbf{Atom} & \textbf{Wyckoff Position} & \textbf{x} & \textbf{y} & \textbf{z} & \textbf{Occupancy} \\
		\hline
		Sr & 4e & 0.0 & 0.5 & 0.25 & 1.00 \\
		Fe & 4a & 0.0 & 0.0 & 0.0 & 0.50 \\
		Ru & 4a & 0.0 & 0.0 & 0.0 & 0.50 \\
		O  & 4e & 0.0 & 0.97(7) & 0.25 & 1.0 \\
		O  & 8f & 0.23(6) & 0.76(2) & 0.01(1) & 1.0 \\
		\hline
	\end{tabular}
	\label{param}
\end{table}
}

SEM was employed to examine the particle morphology of the synthesized material. As shown in figure~\ref{str}(c), the particles exhibit an average size in the micron range, confirming the bulk nature of the compound.  

Figures~\ref{str}(d) and~\ref{str}(e) display the HRTEM image and its corresponding inverse fast Fourier transform (IFFT), respectively. These images clearly demonstrate the high crystallinity of the sample. To verify the elemental composition, EDX analysis was performed, as presented in figure~\ref{str}(f). The EDX spectrum reveals the atomic percentages of the constituent elements—21.2\% Sr, 10.7\% Fe, 10.5\% Ru, and 57.6\% O—which are in excellent agreement with the nominal stoichiometry of SFRO. The absence of any extraneous peaks confirms the chemical purity of the compound.  

The inset of figure~\ref{str}(f) displays a Scanning Transmission Electron Microscopy–High-Angle Annular Dark-Field (STEM–HAADF) image of the region (marked by a red square) where EDX mapping was conducted. Elemental mapping of an individual particle, presented in figure~\ref{str}(g), along with the corresponding elemental distribution maps shown in figures~\ref{str}(h)–\ref{str}(m), demonstrates a uniform spatial distribution of all constituent elements across the particle.

\subsection{Magnetic Properties and Griffiths Phase}

\begin{figure*}
	\centering
	\includegraphics[width=.9\linewidth]{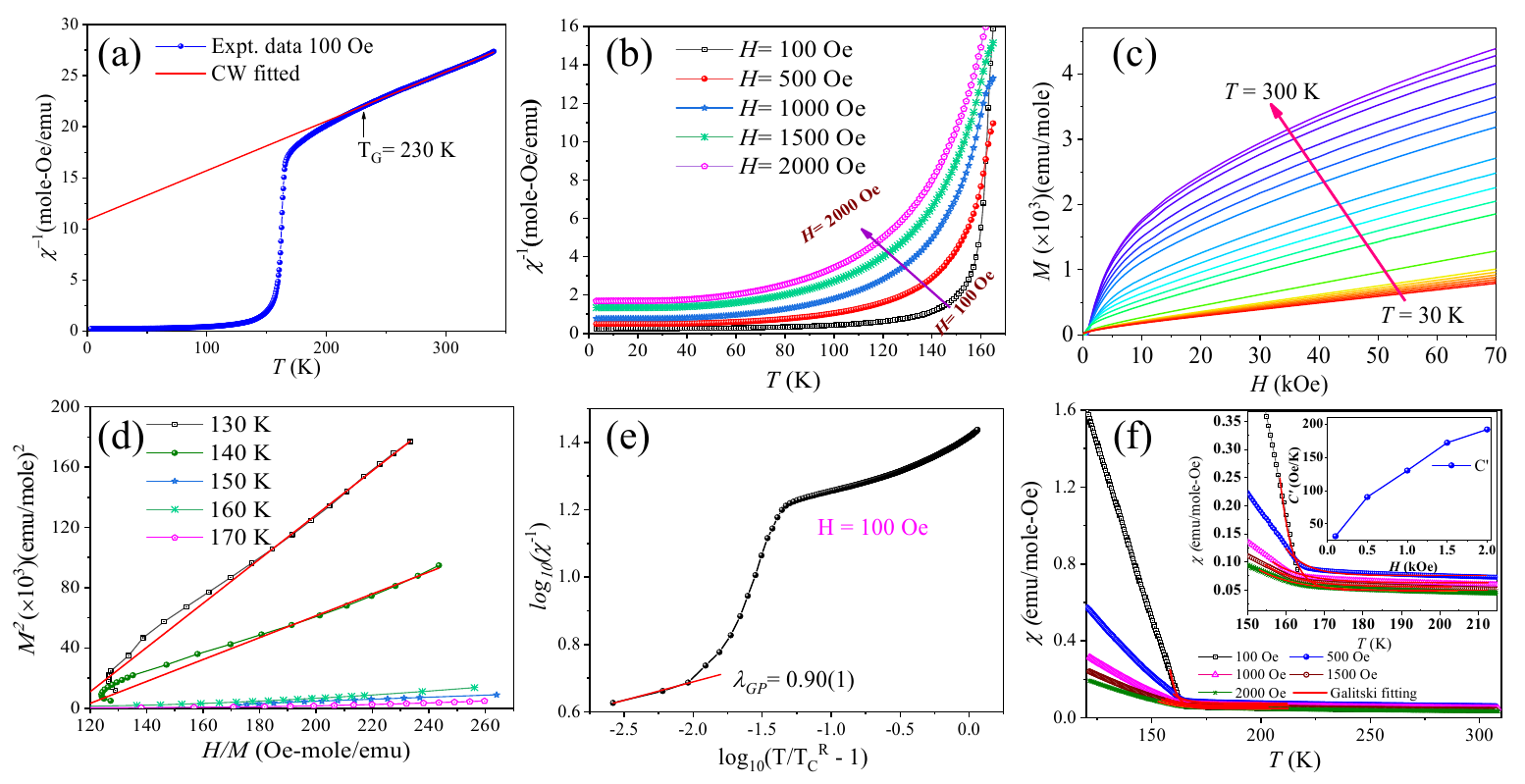}
	\caption{(a) Temperature-dependent inverse susceptibility ($1/\chi$) curve for the SFRO compound measured under an applied magnetic field of 100~Oe. The solid red line represents the Curie–Weiss (CW) fit. (b) $1/\chi$ under external magnetic fields of 100~Oe, 500~Oe, 1500~Oe, and 2000~Oe. The solid lines serve as guides to the eye, highlighting deviations from CW behavior. (c) Isothermal magnetization curves recorded between 30~K and 300~K for the SFRO compound. (d) Arrott plots ($H/M$ vs $M^2$) above the transition temperature. Solid lines denote linear extrapolations from the high-field region. (e) Double-logarithmic plot of inverse susceptibility data against the reduced temperature, with solid red lines indicating linear fits in the Griffiths phase (GP) region. (f) Galitski model fitting for magnetic fields ranging from 100~Oe to 2000~Oe. The inset shows a magnified view of the GP region, while the inset of the inset displays the variation of the estimated constant $C'$ with the applied magnetic field.}
	\label{mag}
\end{figure*}

To probe the magnetic interactions in SFRO, the temperature-dependent magnetization $M(T)$ was measured under both zero-field-cooled (ZFC) and field-cooled (FC) protocols over the range of 3–340~K at different applied magnetic fields. Figure~\ref{mag}(a) shows the temperature-dependent inverse susceptibility curve, $\chi^{-1}(T)$, measured under an applied field of 100~Oe. A small downward deviation from the Curie–Weiss (CW) behavior appears below 230~K, while above this temperature, $\chi^{-1}(T)$ follows a typical CW trend. The CW fitting yields a Weiss temperature of $\theta_W = -225$~K, indicating dominant AFM interactions in SFRO.

Figure~\ref{mag}(b) illustrates the variation of inverse susceptibility under different magnetic fields (100–2000~Oe). A pronounced downward deviation is observed well above the magnetic ordering temperature, signifying the formation of short-range FM clusters embedded within the PM matrix~\cite{Ghosh2025, Dutta2025, Saha2025, Phan2022, Pal2019B}. Such behavior is characteristic of the GP, as described by the theories of Griffiths and Bray~\cite{Griffiths1969, bray1982}, where magnetic inhomogeneities lead to non-Curie–Weiss behavior in the PM regime.

To determine whether spontaneous magnetization or long-range order exists above 155~K, Arrott plots ($H/M$ vs $M^2$), derived from the isothermal magnetization data [figure~\ref{mag}(c)], are shown in figure~\ref{mag}(d). The high-field regions were linearly extrapolated, and the intercepts on the negative $M^2$ axis confirm the absence of spontaneous magnetization and long-range FM ordering above 155~K.

{\color{black}
	To confirm the Griffiths phase below \(T_G \approx 230\) K, the inverse susceptibility data were analyzed using the modified Curie--Weiss relation
	\[
	\chi^{-1}(T) \propto (T-T_C^R)^{1-\lambda},
	\]
	where \(\lambda \to 0\) in the paramagnetic regime and \(0 < \lambda < 1\) in the Griffiths phase. The Griffiths phase fitting was performed in the range of 155~K--230~K. The reference temperature \(T_C^R\) was determined self-consistently by varying trial values in the high-temperature paramagnetic region and selecting the value for which the fitted exponent in that region became statistically indistinguishable from zero, while also giving the best linearity of the fit. Using this \(T_C^R\), the Griffiths exponent \(\lambda_{\mathrm{GP}}\) was extracted in the range \(T_N < T < T_G\). The fitting was repeated for nearby \(T_C^R\) values and slightly different temperature windows to estimate the uncertainty and confirm the stability of the obtained \(\lambda_{\mathrm{GP}}\). Since \(\lambda\) values close to 1 lie near the upper limit of validity of this phenomenological description, they are interpreted cautiously as indicating a very strong Griffiths-like singularity \cite{Jiang2008,Banik2018Ev, Saha2022,bhatti2019magnetic,pramanik2010griffiths}.
}

{\color{black}
	Figure~\ref{mag}(e) presents the double-logarithmic plot of $\log_{10}\chi^{-1}$ versus $\log_{10}(T/T_{\mathrm{CR}} - 1)$, from which the Griffiths exponent $\lambda_{\mathrm{GP}}$ was extracted. The fitted value $\lambda_{\mathrm{GP}} \approx 0.90(1)$ indicates a very strong Griffiths-like deviation from ideal Curie--Weiss behavior ~\cite{Karmakar2013, Nasir2020, Saha2022}. However, since $\lambda$ values close to 1 lie near the upper bound of applicability of the modified Curie--Weiss description, such values should be interpreted with caution. Therefore, the large $\lambda_{\mathrm{GP}}$ obtained here is taken as indicative of a strong Griffiths-like response within the fitting framework, but not as an isolated proof by itself.
}

{\color{black}
To further examine the field dependence of the GP regime, the \(\chi(T)\) data measured under different magnetic fields (100--2000~Oe) were analyzed using the phenomenological expression proposed by Galitski \textit{et al.}~\cite{Galitski2004}. Within this framework, the magnetization in the GP regime follows
\begin{equation}
	M(T, H) \propto \exp[-C'(T/H)],
\end{equation}
where \(C' = k_B / \mu_{\text{eff}}\), with \(k_B\) being the Boltzmann constant and \(\mu_{\text{eff}}\) representing an effective cluster-response scale. The fits, shown in figure~\ref{mag}(f), reproduce the experimental \(\chi(T)\) data reasonably well within the GP region. The inset of figure~\ref{mag}(f) shows that the fitted \(C'\) values increase with the applied magnetic field, indicating a progressive suppression of the anomalous Griffiths-like susceptibility contribution. Accordingly, the corresponding decrease in \(\mu_{\text{eff}}\) should be interpreted as a reduction in the effective fluctuating cluster response captured by the model, rather than as a direct decrease in the intrinsic moment of individual FM clusters. Galitski expression is employed here only in a phenomenological sense, as a supporting description of disorder-induced magnetic inhomogeneity, and not as a unique microscopic model for SFRO ~\cite{Dutta2025, Ghosh2025}.
}

\subsection{Electrical Transport}

\begin{figure}[h!]
	\centering
	\includegraphics[width=0.99\columnwidth]{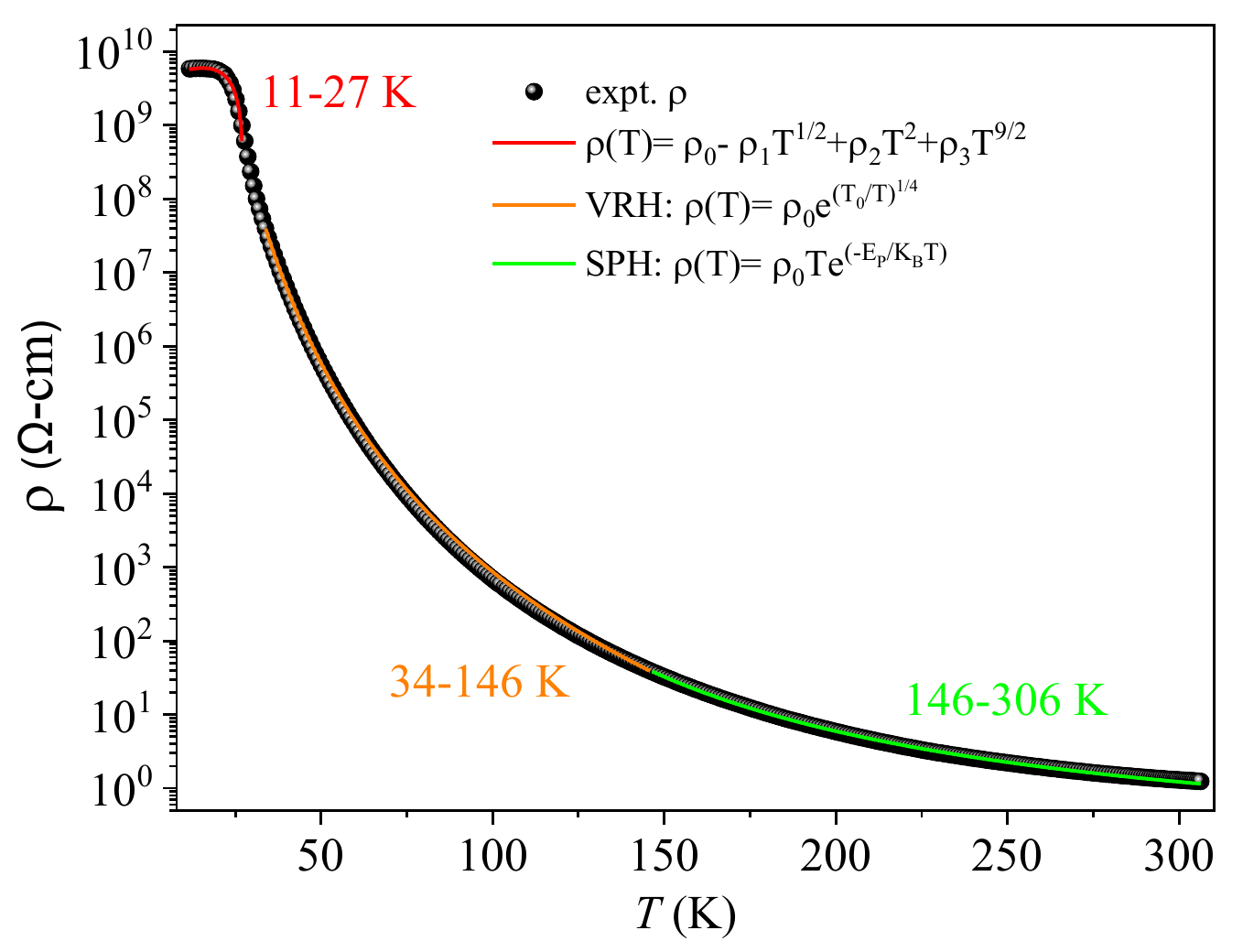}
	\caption{Temperature dependence of electrical resistivity for the SFRO compound, along with the corresponding model fits in different temperature regimes. The green, orange, and red lines represent the best fits obtained using the SPH, VRH, and low-temperature polynomial (LT) models, respectively. The corresponding fitting equations are also indicated.}
	\label{trans}
\end{figure}

The temperature dependence of electrical resistivity for the SFRO compound in the absence of an external magnetic field is presented in figure~\ref{trans}. The resistivity data were recorded during the warming cycle. The temperature-dependent resistivity exhibits a predominantly nonmetallic behavior upon cooling, with a resistivity extremum observed near $T \sim 27$~K.\\

To gain deeper insight into the charge transport mechanisms across various temperature regimes, we analyzed the resistivity data using several established theoretical models. For this purpose, the entire temperature range was divided into three distinct regimes: a high-temperature regime (146–306~K), a low-temperature regime (34–146~K), and a very-low-temperature regime (11–28~K) \cite{Ghosh2024trns}.\\

\begin{figure}[h!]
	\centering
	\includegraphics[width=0.99\columnwidth]{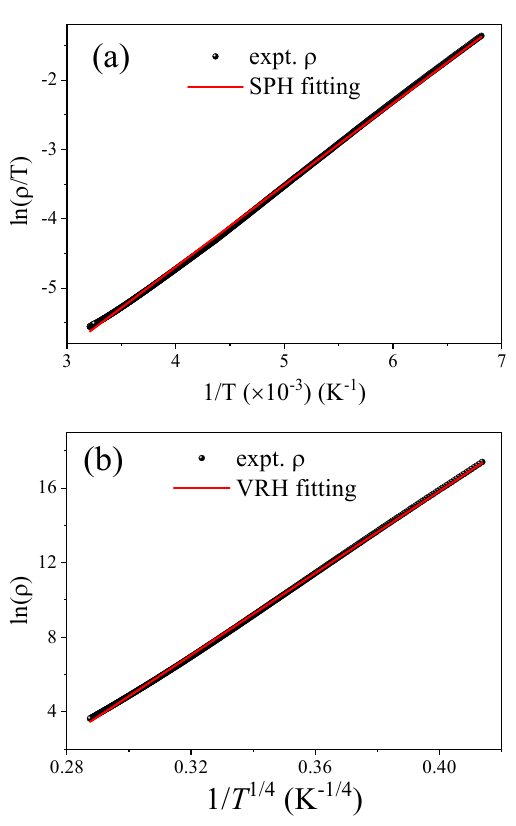}
	\caption{ (a) Small-polaron hopping (SPH) model: variation of $\ln(\rho/T)$ with $1/T$ under zero magnetic field, where the red line represents the best fit to the SPH model; (b) Variable-range hopping (VRH) model: variation of $\ln\rho$ with $T^{-1/4}$ for the SFRO compound in zero magnetic field, with the red line showing the best fit to the VRH model.}
	\label{VRH}
\end{figure}

\noindent

To elucidate the conduction process in the high-temperature region, several models were tested to assess their applicability. Among them, the simple band-gap (Arrhenius) model, characterized by the activation energy $E_A$, is given by~\cite{Sagar2012}
\[
\rho = \rho_0 \exp\left(\frac{E_A}{k_B T}\right),
\]
which predicts a linear relationship in a $\log \rho$ versus $1/T$ plot~\cite{Xiong1995}. However, no such linearity was observed in our data, suggesting that the band-gap model is inadequate for describing the transport in SFRO.  

To interpret the resistivity behavior, we employed the small-polaron hopping (SPH) model~\cite{Mazumdar2021, Amara2017, Modi2017}, which is typically observed in oxides and semiconducting oxide glasses, particularly those containing transition-metal ions (TMI) with mixed valence states. We have fitted this model in the temperature range of 146~K--306~K. In this mechanism, an electron is first thermally excited to an intermediate state, followed by a phonon-assisted hop to a neighboring site~\cite{Sagar2012}. The resistivity in the SPH framework follows~\cite{Banerjee2001, Mazumdar2021}
\begin{equation}
	\rho = A T \exp\left(\frac{E_{P}}{k_{B}T}\right),
\end{equation}
where
\begin{equation}
	A = \frac{k_{B}}{\nu_{ph} N e^{2} R^{2} C (1 - C)} \exp(2\alpha R),
\end{equation}
and $E_{P}$ is the activation energy for polaron hopping. Here, $N$ is the number of ion sites per unit volume, $R = (1/N)^{1/3}$ the average hopping distance, $C$ the fraction of occupied sites, $\alpha$ the decay constant of the electronic wave function, and $\nu_{ph} = k_{B}\theta_{D}/h$ is the phonon frequency, with $\theta_{D}$ being the Debye temperature. From the slope of the $\ln(\rho/T)$ versus $1/T$ plot [figure~\ref{VRH}(b)], the activation energy was estimated to be $E_{P} = 101.75$~meV.\\

With decreasing temperature, the SPH model no longer fits the resistivity data adequately, as the available thermal energy becomes insufficient to overcome the potential barriers for nearest-neighbor hopping. In this regime, loosely bound charge carriers (large polarons) undergo phonon-assisted hopping to energetically favorable distant sites~\cite{Sagar2012, Mazumdar2021}. The corresponding resistivity is described by the variable-range hopping (VRH) model~\cite{Xiong1995, Modi2017, Mazumdar2021}:
\begin{equation}
	\rho = \rho_{0} \exp{\left( \frac{T_{0}}{T} \right)^{1/4}}, 
	\label{eq:VRH1}
\end{equation}
where $\rho_{0}$ is the residual resistivity, and $T_{0}$ is a characteristic temperature defined as
\begin{equation}
	T_{0} = \frac{24}{\pi N(E_{F}) k_{B} \xi^{3}},
	\label{eq:VRH2}
\end{equation}
with $N(E_{F})$ denoting the density of states at the Fermi level and $\xi$ the localization length~\cite{Banerjee2001}.  
The most probable hopping distance, $R_{h}(T)$, and hopping energy, $W_{h}(T)$, are given by
\begin{equation}
	R_{h}(T) = \frac{3}{8} \xi \left( \frac{T_{0}}{T} \right)^{1/4}, \qquad 
	W_{h}(T) = \frac{1}{4} k_{B} T^{3/4} (T_{0})^{1/4}.
\end{equation}

{\color{black}
Figure~\ref{VRH}(a) shows the $\ln \rho$ versus $T^{-1/4}$ plot fitted using the VRH model in the temperature range 34--146~K under zero magnetic field. From the slope, $T_{0}$ was estimated to be $1.46 \times 10^{8}$~K, corresponding to $N(E_{F}) = 1.26 \times 10^{18}$~eV$^{-1}$cm$^{-3}$. 
{\color{black}
	The hopping length and hopping energy are temperature-dependent quantities within the VRH formalism and can therefore be meaningfully discussed only within the fitted VRH regime of 34--146 K, as described by Eq.~(6).
}\\
 We emphasize that these values are approximate, model-dependent estimates, meaningful only within the fitted VRH temperature range, and should not be interpreted as precise intrinsic parameters of the material.}\\

\noindent 
{\color{black}
The emergence of anomalous behavior below \(T_\mathrm{N} \approx 27\) K is evident from the \(\rho(T)\) curve. To describe this region, the resistivity was fitted using the empirical polynomial expression
\begin{equation}
	\rho(T) = \rho_{0} - \rho_{1}T^{1/2} + \rho_{2}T^{2} + \rho_{3}T^{9/2}.
\end{equation}
This expression provides a satisfactory numerical description of the resistivity. However, since SFRO remains highly insulating in this regime, with resistivity of the order of \(10^{10}~\Omega\cdot\text{cm}\), the above form is used here only as an empirical fitting function and no specific microscopic interpretation is assigned to the individual terms.

From the fitting, the parameters were obtained as 
\(\rho_{0} = 1.29 \times 10^{10}~\Omega\cdot\text{cm}\), 
\(\rho_{1} = 2.97 \times 10^{9}~\Omega\cdot\text{cm}\cdot \text{K}^{-1/2}\), 
\(\rho_{2} = 2.46 \times 10^{7}~\Omega\cdot\text{cm}\cdot \text{K}^{-2}\), and 
\(\rho_{3} = -5.66 \times 10^{3}~\Omega\cdot\text{cm}\cdot \text{K}^{-9/2}\). 
These coefficients should therefore be regarded only as fitting parameters that parameterize the observed resistivity trend in the low-temperature region. The fit is employed here solely to capture the phenomenological behavior associated with the resistivity minimum near 27 K, rather than to identify distinct metallic scattering contributions.
}

Overall, the temperature-dependent resistivity of SFRO can be comprehensively described using three distinct conduction mechanisms: SPH in the high-temperature regime ($T > 146$~K), VRH in the intermediate regime ($34~\text{K} < T < 146~\text{K}$), and an anomalous behavior at low temperatures ($T < 27$~K). These findings demonstrate a clear evolution of the charge transport mechanism across temperature scales, reflecting the complex interplay among lattice disorder, polaronic effects, and magnetic correlations in this 3d–4d hybrid double perovskite.

	\section{Conclusion}
In conclusion, the investigation of Sr$_{2}$FeRuO$_{6}$ provides a comprehensive understanding of the intricate correlation between inherent structural disorder and the emergent magnetic and charge-transport phenomena in 3d--4d hybrid double perovskites. The compound crystallizes in a monoclinic structure with pronounced ASD, which serves as the key parameter governing its unconventional physical properties.  

Magnetically, SFRO is dominated by strong AFM exchange interactions ($\theta_{W} = -225$~K), while structural frustration induces the formation of a robust Griffiths Phase (GP) persisting up to $T_{G} = 230$~K. The high singularity exponent ($\lambda_{GP} \approx 0.90(1)$) quantitatively reflects extreme magnetic inhomogeneity arising from short-range ferromagnetic clusters. The pronounced field dependence of the GP further indicates potential tunability of these localized magnetic correlations.  

Electrically, the structural disorder imposes severe carrier localization, leading to high resistivity and a distinct temperature-dependent crossover in transport mechanisms. Conduction proceeds via SPH at high temperatures ($E_{P} = 101.75$~meV) and transitions to VRH at intermediate temperatures, characterized by a large characteristic temperature ($T_{0} = 1.46 \times 10^{8}$~K), indicative of strong localization. At low temperatures, transport is predominantly influenced by extrinsic factors such as structural defects and domain boundary scattering.

{\color{black}
	Overall, SFRO emerges as a system with substantial structural and magnetic inhomogeneity, in which Griffiths-phase-like spin correlations and highly localized charge transport are both observed.}
	These findings on Sr$_{2}$FeRuO$_{6}$ offer crucial insights into how lattice imperfections and competing interactions in perovskites promote the emergence of Griffiths-phase-like transport phenomena.

\section*{Supporting Information}

he Supplementary Information provides a comprehensive analysis of magnetization and relaxation dynamics, detailed Griffiths phase analysis methods, current–voltage characteristics, and XPS measurements.

\section*{Acknowledgments}

They also gratefully acknowledge Rui Wu and the South China University of Technology, China, for the fellowship and other assistance. We also acknowledge Suman Dey and Biswarup Satpati for their support with the TEM imaging. The authors would like to thank the Saha Institute of Nuclear Physics, DAE, India, for providing support.

	\section*{Data Availability Statement}

The data that support the findings of this study are available from the corresponding author upon reasonable request.
\\

	\section*{Reference}

\bibliographystyle{unsrt}
\bibliography{ref}

\end{document}